\documentstyle[12pt]{article}

\begin{document}

\begin{center}
{\Large {\bf
The Gerasimov-Drell-Hearn Sum Rule}}
\\
{\Large {\bf
and  Current Algebras}}
\end{center}
\vskip 0.20in
\begin{center}
Lay Nam Chang$^a$\footnote{Presently at
Physics Division,
National Science Foundation, Arlington, Virginia 22230,
under a contract with
National BioSystems, Rockville, Maryland 20852.}
\quad Yigao Liang$^{b}$\footnote{Present address: Integrated
Decisions and Systems, 8500 Normandale Lake Boulevard, Suite 1840,
Bloomington, MN 55437}
\quad Ron L.\ Workman$^{a}$ \\
$^{a}\;$ Department of Physics and Institute for High Energy Physics \\
Virginia Polytechnic Institute and State University \\
Blacksburg, Virginia 24061-0435 \\
$^{b}\;$ Department of Physics and Astronomy \\
University of Rochester \\
Rochester, New York 14627
\end{center}
\vskip 0.20in
\begin{abstract}
The status of the Gerasimov-Drell-Hearn sum rules for
polarized inclusive photo-production on nucleons is reviewed.
It is shown that results from currently available data compare
favorably with an estimate based on an extended current algebra.
Implications for integrals of spin-dependent structure functions
are also briefly discussed.
\end{abstract}
\vfil
\eject

The Gerasimov-Drell-Hearn (GDH)
sum rule\cite{gdh} has recently regained
the attention of both theorists and experimentalists.  This renewed
interest is partly due to the construction of CEBAF.  There now exist
proposals for measurements which will check both this sum rule and
its generalization to non-zero values of $Q^2$.  On the theoretical
side there are connections\cite{an} to the Bjorken sum rule and the
EMC measurements\cite{emc}.
As discrepancies exist in the predicted low-$Q^2$ behavior,
it will be useful to better understand the physics at $Q^2=0$.

The original GDH sum rule, which actually
results from a superconvergence relation
for the spin-flip amplitude in Compton scattering, states that
\begin{equation}
{ {2 \pi^2 \alpha} \over {M^2} } (\kappa_{p,n})^2 =
\int\limits_{\omega_0}^{\infty} { [{\sigma_{3/2} (\omega) -
\sigma_{1/2} (\omega)]_{p,n} } \over {\omega} } d \omega,
\end{equation}
where, $\kappa_p$($\kappa_n$) is the proton(neutron) anomalous
magnetic moment, $\omega$ is the laboratory photon energy,
$M$ is the nucleon mass, and $\alpha$ is the fine structure constant.
The left-hand-side represents the single nucleon contribution
to the spin-flip amplitude, while the right-hand-side involves
an integration over the difference of
helicity 3/2 and helicity 1/2 $\gamma p$ total cross sections.

Karliner\cite{kar} examined Eq.(1) decomposed into isoscalar and
isovector components.  The isovector sum rule was found to be well-satisfied.
The isoscalar component, which is predicted to be small, was found to be
small but difficult to determine precisely. The isovector-isoscalar sum
rule provided the most unexpected result.  Karliner's estimate of the
integral of cross sections was positive; the value from the
anomalous magnetic moments is negative. These conclusions have been
reproduced in a more recent study\cite{work}.

The problem with the isovector-isoscalar sum rule
was also noticed by Fox and Freedman
in their extensive study of Compton scattering sum rules\cite{fox}.
Earlier, Abarbanel and Goldberger\cite{ab} had noted that the GDH sum rule
would be modified if there were a fixed pole at $J=1$ in the angular
momentum plane. Fox and Freedman suggested that this discrepancy might
be evidence for such a pole.

Such fixed poles can be attributed generally to non-trivial
terms in the associated current commutators\cite{ca,ks}.
In this context, the superconvergence relation giving rise to the GDH
sum rule implies that electric charge densities commute with each other.
Within the standard model, electric charge densities are bilinear
in fermion fields, and a naive application of canonical
anti-commutation relations substantiates this conclusion.

However, a more careful study carried out recently
shows that the commutator actually
does have additional terms\cite{cla}.
Indeed, these terms have
quantum numbers that can give rise to modifications of the
GDH sum rule.
In this letter,
we estimate the contribution of these
terms, and compare the result with that of the latest phenomenological
analysis\cite{work}.

The crucial point of ref.\cite{cla} is that the
vacuum expectation value of the triple commutator of charge densities
 generated by quark fields cannot vanish,
and is proportional to the number of colors carried by these fields.
Furthermore, its structure
implies that
the charge density commutator
$ [ V^{a} ({\bf x}) , V^{ b} ({\bf y}) ] $ must have a term
symmetric in the $a,b$ indices.   This is in addition to
the usual anti-symmetric term,
$i f^{abc} V^{ c} ({\bf x})
\delta^{3} ({\bf x -y}) $.
Finally,
the extra term must be of the form
$   d^{abc} [ \nabla \times {\bf b}^{c}_{A} ({\bf x}) ] \cdot \nabla
\delta^{3} ({\bf x-y}) $ in order that the
charge densities transform properly under
global flavor transformations.   Here, ${\bf b}^{c}_{A}$ is
an axial-vector operator whose precise form is dependent upon
specific dynamics.

The minimal
(infinite dimensional) Lie algebra that
reproduces the correct vacuum expectation value of the
triple commutators of the vector and axial-vector
charge densities is given by the following relations\cite{cla}:
\begin{eqnarray}  \label{ca}
 [ V^{a} ({\bf x}) , V^{ b} ({\bf y}) ] & = &
i f^{abc} V^{ c} ({\bf x})
\delta^{3} ({\bf x -y}) + {i \over 48 \pi^{2}  } d^{abc}
[ \nabla \times {\bf b}^{c}_{A} ({\bf x}) ] \cdot \nabla
\delta^{3} ({\bf x-y}),  \cr
 [ V^{a} ({\bf x}) , A^{ b} ({\bf y}) ] & = &
i f^{abc} A^{ c} ({\bf x})
\delta^{3} ({\bf x -y}) + {i  \over 48 \pi^{2}  } d^{abc}
[ \nabla \times {\bf b}^{c}_{V} ({\bf x}) ] \cdot \nabla
\delta^{3} ({\bf x-y}),  \cr
 [ A^{a} ({\bf x}) , A^{ b} ({\bf y}) ] & = &
i f^{abc} V^{ c} ({\bf x})
\delta^{3} ({\bf x -y}) + {i  \over 48 \pi^{2}  } d^{abc}
[ \nabla \times {\bf b}^{c}_{A} ({\bf x}) ] \cdot \nabla
\delta^{3} ({\bf x-y}) , \cr
 [ V^{a} ({\bf x}) , {\bf b}_{V}^{ b} ({\bf y}) ] & = &
i f^{abc} {\bf b}_{V}^{ c} ({\bf x})
\delta^{3} ({\bf x -y}) + 2i N_{c} \delta^{ab}
\nabla \delta^{3} ({\bf x-y}),  \cr
 [ A^{a} ({\bf x}) , {\bf b}_{A}^{ b} ({\bf y}) ] & = &
i f^{abc} {\bf b}_{V}^{ c} ({\bf x})
\delta^{3} ({\bf x -y}) + 2i N_{c} \delta^{ab}
\nabla \delta^{3} ({\bf x-y}),  \cr
 [ A^{a} ({\bf x}) , {\bf b}_{V}^{ b} ({\bf y}) ] & = &
i f^{abc} {\bf b}_{A}^{ c} ({\bf x})
\delta^{3} ({\bf x -y}) , \cr
 [ V^{a} ({\bf x}) , {\bf b}_{A}^{ b} ({\bf y}) ] & = &
i f^{abc} {\bf b}_{A}^{ c} ({\bf x})
\delta^{3} ({\bf x -y}) .
\end{eqnarray}
The generators of the algebra are the charge densities,
the {\bf b} terms and a central charge $N_c$,
which is equal to the number of colors for the quarks.
The number of colors did not appear in the older current
algebras.   Note that despite the appearance of an extension term
in the charge density commutator, the
electromagnetic current
is not anomalous as a gauge current\cite{let}.

However, upon application of standard techniques\cite{ks,adler},
the extension term does
yield a modified form of the GDH sum rule:
\begin{equation}  \label{newgdh}
{ {2 \pi^2 \alpha} \over {M^2} } (\kappa_{p,n})^2 + S_{p,n} =
\int\limits_{\omega_0}^{\infty} { {[\sigma_{3/2} (\omega) - \sigma_{1/2}
(\omega)]_{p,n}} \over {\omega} } d \omega.
\end{equation}
Here the subscripts $p,n$ label the proton and neutron sum rules. The
additional contribution, $S_{p,n}$, is due to the symmetric term.
The discrepancy found in the traditional sum rule, if
it persists, is an indication that the term  $S_{p,n}$ does not
vanish.  In the following, we describe a scenario in which
an $S_{p,n}$ term appears naturally with a value that
accounts for the apparent discrepancy in the sum rule.

We begin by observing that the operators ${\bf b}_{V,A}$ defined
by the charge density commutators in Eq.(\ref{ca})
share the same quantum numbers as
the conventional vector and axial-vector currents.
However, except in singular limits\cite{cla}, they
cannot be directly proportional as operators to these currents.
The
Schwinger terms in the commutators of the charge densities and
the spatial current densities are infinite in QCD,
while the corresponding quantity in Eq.(\ref{ca}) takes on the
finite value $2N_c$.    Nevertheless, certain classes of matrix elements
of ${\bf b}_{V,A}$ may well be proportional to those of the currents.   In
the following, we will examine this possibility for the class of
single-particle matrix elements
of this operator.
Specifically, we will assume that 
single-particle matrix elements of ${\bf b}_{A,V}$
are well-approximated by
contributions from the nucleon,
$\pi$, $a_1$ and $\rho$,  and that their
relative strengths among these states are the same as the corresponding ones
for the weak currents ${\bf A}$ and ${\bf V}$.
Such a hypothesis is tantamount to supposing that one may realize
the extended algebra by local fields with these quantum numbers within
a specific field-theoretic context.
We symbolically write the relevant matrix elements as
${\bf b}_A = K_A {\bf A}_L$,
${\bf b}_V = K_V {\bf V}_L$, where the subscript $L$ indicates
that only the couplings to the low-lying states are included.

Within such a framework,
 the {\bf b} operators are effectively decoupled
from the high energy states, and  two point functions
of the form $ < b_A A > $ and $ < b_V V > $ are finite,
consistent with Eq.(\ref{ca}).
Assuming the validity of the KSRF relation\cite{ksrf},
the spectral sum rule from $ < b_A A > $ equals
$2 f^2_\pi K_A $,
half coming from the pion and the other half from $a_1$\cite{weinberg}.
The spectral  sum rule from   $ < b_V V > $, in the same approximation,
is given by
$2 f^2_\pi K_V $.
Comparing the spectral sums with the Schwinger terms in Eq.(\ref{ca}),
we have, for $N_c = 3$,
\begin{equation} \label{K}
K_A=K_V=3/ f_{\pi}^{2} .
\end{equation}

We may compare this result with what can be expected within the
context of an effective field theory for QCD at low energies,
such as the non-linear sigma model\cite{sw,wz}.
For simplicity, we restrict our attention to the minimal model involving
only pions, and
consider the 2-flavor case\cite{ylunpub} (for a discussion of the
case with more than two flavors,
see \cite{kr}).  Within this $SU(2)$ framework there is no canonical
isosinglet current, and the associated electromagnetic current
cannot describe properties of the nucleon.
As pointed out in\cite{witten2}, for purposes
of studying these properties, one
should include a topological current which is to be identified
as the isosinglet
baryon number current.
The full electromagnetic current relevant for both the nucleon and
the pion
is the sum of
the canonical isotriplet current and this topological current.
Using canonical commutation relations, we can check that with
this augmentation
the algebra of Eq.(\ref{ca}) is indeed reproduced.
More specifically, the commutator of the electromagnetic charge
densities
has an extension term that is purely isotriplet, and
takes the
form ${\bf b}_A = K {\bf A}$, where
${\bf A}$ is the {\it exact} isotriplet
axial-vector current in the model.

The precise value of $K$ is dependent upon the dynamics relevant
to the energy scale under consideration;  it is given in
the minimal non-linear sigma model by
$K_A =24 / f_{\pi}^{2}$,
which is much larger than
the estimate quoted in Eq.(\ref{K}).   The difference is
primarily due
to the absence of the $a_1$ and $\rho$ mesons in this model.
The KSFR relation assumed in our estimate above implies equal
contributions of the pion and the $a_1$ to the Schwinger term
coefficient of the extension current density in Eq.(\ref{ca}).
Through the Weinberg sum rule, their sum gives the
contribution of the $\rho$ to the Schwinger term coefficient.
Since this coefficient in Eq.(\ref{ca}) is fixed
to be the number of colors $N_c$, the
proportionality factor must be correspondingly larger to make up
for the missing contributions from the spin-1 mesons.

We will be using Eq.(\ref{K}) in the subsequent analysis.
The contributions $S_{p,n}$
to the extension term then have the values:
\begin{eqnarray} \label{s}
     S_{(p,n)} &=&    \alpha {(g_{A})_{(p,n)}  K_{A} \over 18}  \\
(g_{A})_{(p,n)} &=& {1\over \sqrt{3}}G_{A}^{(8)} \pm G_A^{(3)}
\end{eqnarray}
when we substitute for the flavor $SU(3)$
$d$-symbols.
 The superscripts denote the directions
 in this flavor space, with $G_A^{(3)}=1.25$
being the Gamow-Teller $\beta$-decay constant.    The value for $G_A^{(8)}$
is not known directly at this time.   However, if one supposes flavor
$SU(3)$ symmetry, then by fitting measured hyperon decay rates, one
can roughly estimate that
$G_A^{(8)}=0.65$, based upon
an $F/D$
ratio of $0.63$\cite{fd}.   In this work,
we ignore the contribution from the singlet
axial current, since the EMC experiments\cite{emc} suggest that
with the same $F/D$ ratio,
$G_A^{(0)} \sim 0.12$,

We may eliminate all dependence on the poorly determined $G_A^{(8)}$ by
taking the difference of the proton and neutron GDH sum rules:
\begin{equation}  \label{sv}
{ {2 \pi^2 \alpha} \over {M^2} } [(\kappa_p)^2 - (\kappa_n)^2] + S =
\int\limits_{\omega_0}^{\infty} { {[\sigma_{3/2} (\omega) - \sigma_{1/2}
(\omega)]_p - [\sigma_{3/2} (\omega) - \sigma_{1/2}
(\omega)]_n } \over {\omega} } d \omega.
\end{equation}
This form is referred to as the isovector-isoscalar sum
rule\cite{kar,work}.   The quantity $S$ denotes the difference $S_p - S_n$.
The first term on the left-hand-side is actually negative; with
$\kappa_p = 1.793$, and $\kappa_n =-1.913$, it equals
$-28 \mu$b.
The right-hand-side has been estimated in Refs.\cite{kar,work}
to lie between 50$\mu$b and 70$\mu$b.  This
range, however,
clearly underestimates the uncertainty
of this quantity, since
no error has been assigned to contributions coming from the $\gamma p \to
\pi \pi N$ process.  The integral has also been cut off at 2 GeV in the
laboratory photon energy.   Nevertheless, one can see a definite discrepancy
in this component of the GDH sum rule.

The value of the extension term $S$ obtained from
Eqs.(\ref{K}) and (\ref{s}) is,
\begin{equation}
S =  {  \alpha {G_A^{(3)}}  \over 3 F_{\pi}^{2} }.
\end{equation}
With $\alpha = 1/137$, $G_{A}^{(3)} = 1.25$, and $F_{\pi}= 93 MeV$,
we obtain $S$ = 137$\mu b$.
Therefore the left-hand-side of Eq.(\ref{sv}) now has a total value
$- 28 \mu \mbox{b} + 137 \mu \mbox{b} = 109 \mu \mbox{b}$,
which brings Eq.(\ref{sv}) into qualitative agreement
with the recent analyses of Ref.\cite{work}.

Using $SU(3)$ flavor symmetry, we may also estimate the corrections to
the individual proton and neutron GDH sum rules.  The magnetic moment
contribution to the left hand side of the proton sum rule is 205$\mu$b.
Using the result given in Eq.(4) we have
$S_p = 89\mu$b, which increases the total to 294$\mu$b.
The integral over the proton cross-sections is estimated to be
260$\mu$b\cite{work}.   For neutrons, the magnetic
moment contribution is 233$\mu$b,
while $S_n = -48\, \mu$b.  The left-hand-side therefore equals 185$\mu$b,
which is to be compared with the estimate of 190$\mu$b for the integral
Given the uncertainties
involved in these comparisons, the modified GDH sum rules appear to be
reasonably well-satisfied.

The above results involve real photons, with $Q^2 = 0$.
What happens when $Q^2 > 0$,
as in deep inelastic electron and muon scatterings?
Recent analyses of the EMC\cite{emc} data suggest that the integrals over
the moments of the polarized spin structure function, which are in effect
the cross-sections appearing in GDH sum rules for virtual photons, must change
sign as $Q^2 \to 0$ if the original GDH sum rules
without extension modifications
are to be satisfied\cite{an}.   An exception
had been the isovector-isoscalar integral, which is constrained to equal
$-G_A^{(3)}/6$ for large $Q^2$ by the Bjorken sum rule\cite{bj}.   This
sum rule joins onto the isovector-isoscalar GDH sum rule, as
$Q^2 \to 0$, and without the modification $S$ in
Eq.(\ref{sv}), one would have
expected the integral to remain negative.  No sign change
appeared necessary for this integral.  In actual fact, however,
Eq.(\ref{sv}) and recent analyses\cite{work} of available data at $Q^2 = 0$
both suggest that there is a sign change in this case as well.

Why do these sign-flips occur?   From our present perspective, such sign-flips
represent the cross-over from short-distance physics, where perturbative
QCD works, to large-distance non-perturbative physics.   In perturbative
QCD, the extension terms should vanish.
We therefore do not expect any modifications of sum
rules such as the Bjorken sum rule, which are valid for large values
of $Q^2$.
For smaller values of $Q^2$, however, we can approximate large-distance
physics through contributions of low-lying resonances, and try to
saturate the spectral sum rules with these states\cite{weinberg}.
The success of the consequent
predictions suggest that the  {\bf b} operator only has
significant couplings to low-lying state, giving rise to
finite values for the extension terms, and
corrections to the GDH sum rules.  As noted above,
the relevant integrals change sign as a result.

The verification of any sum rule is an evolving process, since all comparisons
must necessarily include cut-offs for the various integrals.
Data from experiments planned at CEBAF, which will probe
intermediate values of $Q^2$,
can only shed more light
on the physics behind the issues presented
in this Letter.

This work was supported in part by the U.S. Department of
Energy Grants DE-FG05-92ER40709, DE-FG05-88ER40454, and DE-FG02-91ER40685.

\end{document}